\documentclass[twocolumn,prl]{revtex4}

\usepackage{amsmath}
\usepackage{amsfonts}            
\usepackage{amssymb}

\usepackage[colorlinks,urlcolor=blue,linkcolor=blue,pagecolor=blue,filecolor=blue,citecolor=blue]{hyperref}

\usepackage{mathrsfs}

\usepackage{eufrak}

\usepackage{verbatim} 

\usepackage{graphicx}
\usepackage{epstopdf}

\usepackage{bm}

\begin{document}

\title{An analytic model of the stereodynamics of rotationally inelastic molecular collisions}

\author{Mikhail Lemeshko} 
\email{mikhail.lemeshko@gmail.com}

\author{Bretislav Friedrich}
\email{brich@fhi-berlin.mpg.de}

\affiliation{%
Fritz-Haber-Institut der Max-Planck-Gesellschaft, Faradayweg 4-6, D-14195 Berlin, Germany
}%

\date{\today}

\begin{abstract}

We develop an analytic model of vector correlations in rotationally inelastic atom-diatom collisions and test it against the much examined Ar--NO ($\mathbf X^2 \Pi$) system. Based on the Fraunhofer scattering of matter waves, the model furnishes complex scattering amplitudes needed to evaluate the polarization moments characterizing the quantum stereodynamics. The analytic polarization moments are found to be in an excellent agreement with experimental results and with close-coupling calculations available at thermal energies. The model reveals that the stereodynamics is governed by diffraction from the repulsive core of the Ar--NO potential, which can be characterized by a single Legendre moment.

\end{abstract}

\maketitle

Observing correlations among the vectors that characterize a collision can disclose all there is to know about \emph{how} the collision proceeds~\cite{VectCorrReviews}. Dudley Herschbach~\cite{HerschbachQuote} likened vector correlations to ``forbidden fruit'' whose ``tasting'' reveals what would otherwise remain hidden. An example he frequently cites is the undoing of the azimuthal averaging about the initial relative velocity vector via a three-vector correlation, which reveals stereodynamical features lost by averaging over the initial distribution of impact parameters. The pioneering work of Herschbach and coworkers~\cite{HerschbachCorrelations} on vector correlations in the domain of molecular collisions spurred an effort to extract the information hidden in molecular dynamics computations, both quasiclassical and quantum, as these contain vector correlations as a default bonus~\cite{QCTquant}. However, even when characterized to the full by vector correlations, the \emph{why} of collision stereodynamics can only be answered as well as the theoretical method applied to treat the collisions allows. In the present work, we extract vector correlations from an analytic model of direct rotationally inelastic atom--diatom collisions, and thereby gain a particularly simple, yet perspicacious insight into their stereodynamics. 

The collision model employed is based on the Fraunhofer scattering of matter waves~\cite{EarlyFraunhofer}, recently extended to treat collisions in fields~\cite{LemFriFraunhofer}. In contrast to classical or semiclassical theories, the Fraunhofer model furnishes complex scattering amplitudes needed to extract the characteristics of vector correlations that reflect the quantum stereodynamics. Owing to its quantum nature, the model accounts for diffraction, interference, and other nonclassical effects. As an example, we deal with inelastic collisions of closed-shell atoms with rotationally polarized symmetric-top-equivalent linear molecules, represented by the much examined Ar--NO ($X^2 \Pi$) system\cite{ArNOstudies,Alexander99}.  The vector correlations obtained from the Fraunhofer model are found to be in an excellent agreement with the results of experiments and close coupling calculations of Wade~\textit{et al.}~\cite{Wade04}. This allows interpreting the collision stereodynamics of the Ar--NO ($X^2 \Pi$) system in terms of the Fraunhofer model.

The Fraunhofer model was described in detail in refs.~\cite{EarlyFraunhofer, LemFriFraunhofer}. It is based on the sudden approximation, which treats the rotational motion as frozen during the collision and thereby allows expressing the inelastic scattering amplitude in terms of the elastic one. The elastic scattering amplitude is, in turn, expressed in terms of the amplitude for Fraunhofer diffraction of matter waves from a sharp-edged, impenetrable obstacle acting in place of the molecular scatterer. At collision energies of hundreds of cm$^{-1}$, consistent with the sudden approximation, the shape of the scatterer is approximated by the repulsive core of the atom--molecule potential, with the attractive part disregarded. The Fraunhofer model renders fully state- and energy-resolved scattering amplitudes and all the quantities that unfold from them in analytic form.

The stereodynamics of an atom--diatom collision is usually described by a set of four vectors: the initial and final relative velocities, $\mathbf{k}$ and $\mathbf{k'}$, and the initial and final rotational angular momenta of the diatomic molecule, $\mathbf{j}$ and $\mathbf{j'}$.  We use the initial and final relative velocities $\mathbf{k}$ and $\mathbf{k'}$ to define the $XZ$ plane of the space-fixed coordinate system, with the initial relative velocity $\mathbf{k}$ pointing along the $Z$ axis. In keeping with the convention of Orr-Ewing and Zare~\cite{OrrEwingZare94}, we characterize the spatial distribution of the angular momenta relative to the $XZ$ plane by the polarization moments $A^{(k)}_{q\pm}$, which arise as coefficients in the expansion of the density operator over the state multipoles~\cite{BlumBook}. Since within the Fraunhofer model the scatterer is two-dimensional, the model can only account for alignment, but not for orientation~\cite{LemFriFraunhofer}. As a result, all polarization moments with odd $k$ or $q$ vanish within the model.


\begin{figure}
\includegraphics[width=8cm]{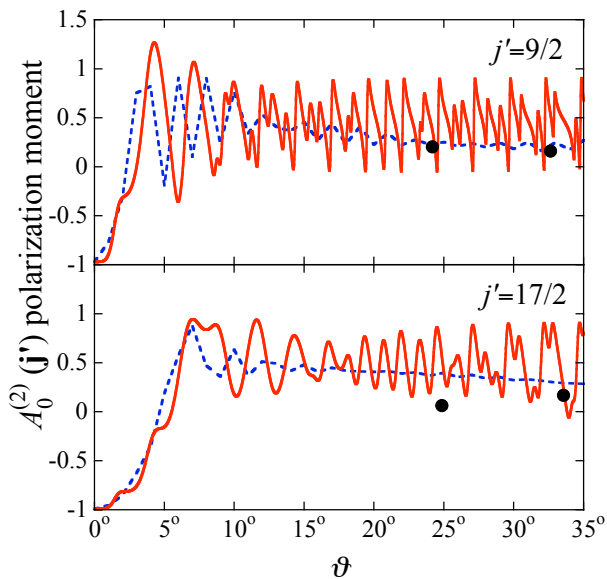}
\caption{\label{fig:A20_moment} Polarization moment $A^{(2)}_0 (\mathbf{j'})$ pertaining to the $\mathbf{k-k'-j'}$ three-vector correlation in Ar--NO~$(j=\Omega=1/2, \to j', \Omega'=1/2)$ collisions at 520$\pm$70~cm$^{-1}$. The analytic results furnished by the Fraunhofer model (red solid line) are compared with the experiment (black dots) and close-coupling calculations (blue dashed line) of ref.~\cite{Wade04}.}
\end{figure}

\begin{figure}
\includegraphics[width=8cm]{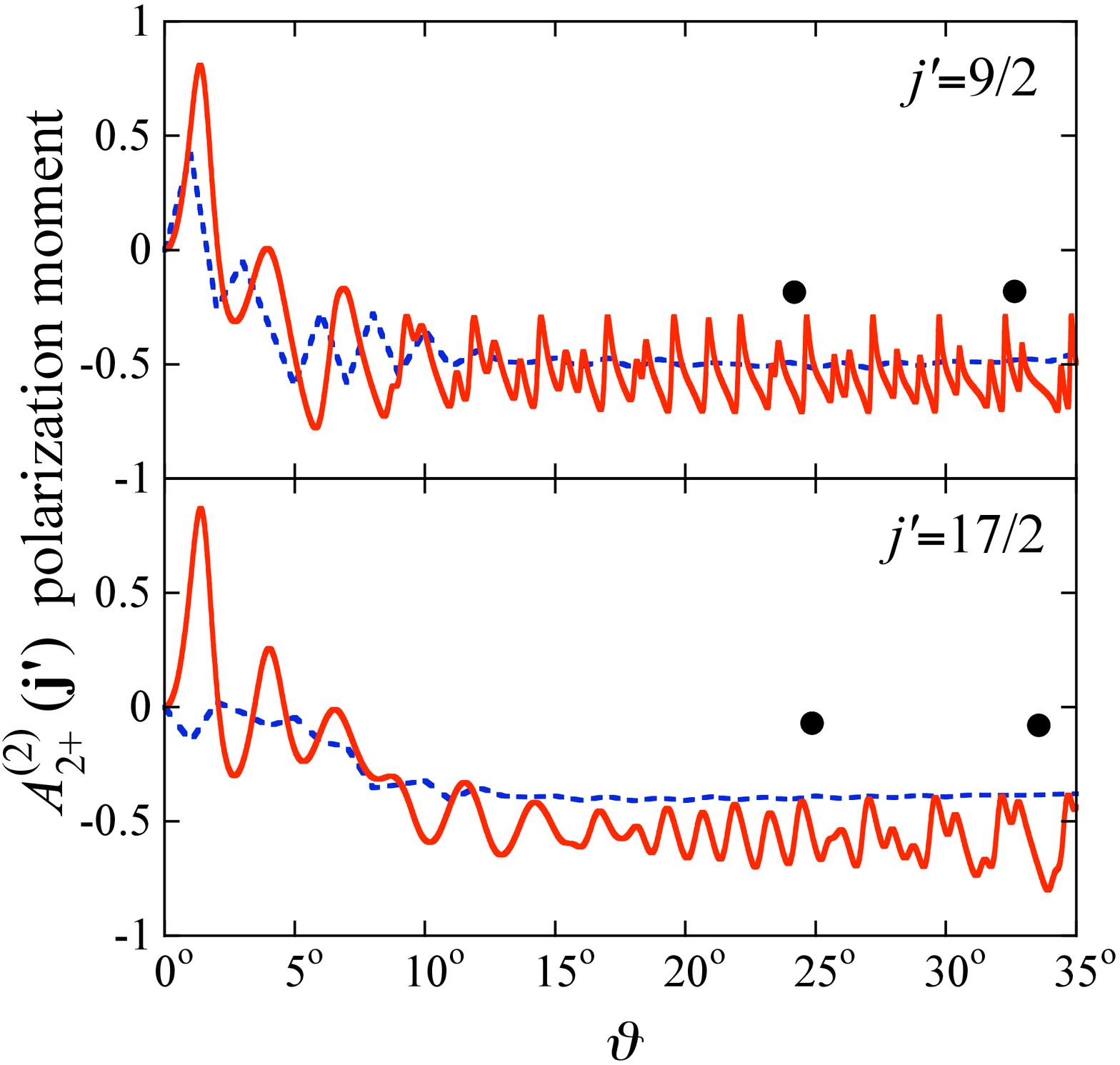}
\caption{\label{fig:A22_moment} Polarization moment $A^{(2)}_{2+} (\mathbf{j'})$ pertatining the $\mathbf{k-k'-j'}$ three-vector correlation in the Ar--NO~$(j=\Omega=1/2, \to j', \Omega'=1/2)$ collision at 520$\pm$70~cm$^{-1}$. The analytic results furnished by the Fraunhofer model (red solid line) are compared with the experiment (black dots) and close-coupling calculations (blue dashed line) of ref.~\cite{Wade04}.}
\end{figure}

For this case study, we chose the $\mathbf{k-k'-j'}$ three-vector correlation in the Ar--NO~($j=\Omega=1/2 \to j', \Omega'=1/2$) collisions, as this can be compared with the results of experiments and close-coupling calculations of Wade~\textit{et al.}~\cite{Wade04}. In addition, we illustrate the scope of the model by treating the $\mathbf{k-j-k'}$ and $\mathbf{k-j-k'-j'}$ correlations in the Ar--NO~($j=\Omega=3/2 \to j', \Omega'=3/2$) scattering which, to date, have not been measured or evaluated. We restrict our considerations to the two lowest rotational channels, $j'=9/2$ and $17/2$, reported in ref.~\cite{Wade04}, and average over the $e/f$ parity states as these have not been resolved in the experiment. We take into account the energy spread of the molecular beams, $E_\text{coll}=520 \pm 70$~cm$^{-1}$, by averaging our results over three collision energies corresponding to the most probable energy and to energies at half-maximum of an essentially Gaussian collision energy distribution. In determining the Ar--NO potential, we rely on the most recent potential energy surface (PES) obtained by Sumiyoshi~\textit{et al.}~\cite{Sumiyoshi07} and make use of only the average potential, $V_\text{sum}$, since the differential and depolarization cross sections are found to be only weakly affected by the difference PES, $V_\text{dif}$~\cite{AoizAlexander}. The PES of ref.~\cite{Sumiyoshi07} comes close to that of Alexander~\cite{Alexander99} and both PES's yield essentially the same polarization moments.
\begin{table}[h]
\centering
\caption{Meaning of the $A^{(2)}_0 (\mathbf{j})$  and $A^{(2)}_{2+}
(\mathbf{j})$ alignment polarization moments. The $Z$ axis points along the
initial relative velocity $\mathbf{k}$. The final relative velocity
$\mathbf{k'}$ lies in the $X>0$ half of the $XZ$ plane. The indicated
ranges of the moments correspond to the high-$j$ limit. The $A^{(2)}_{2-}
(\mathbf{j})$ moment vanishes identically.}
\vspace{0.2cm}
\label{table:moments}
\begin{tabular}{| c | c | c | }
\hline 
\hline
Moment & $A^{(2)}_0 (\mathbf{j})$  & $A^{(2)}_{2+} (\mathbf{j})$ \\[3pt]
\hline
Meaning &   $\mathbf{j}$ along $Z$ & $\mathbf{j}$ along $X$ or $Y$ \\[0.3cm] 
Range & $\mathbf{j} \perp Z$ \hspace{0.01cm} $\to$ \hspace{0.01cm} -1 &   $\mathbf{j} \parallel X$ \hspace{0.01cm} $\to$ \hspace{0.01cm} -1  \\[0.1cm]
  & $\mathbf{j} \parallel Z$ \hspace{0.01cm} $\to$ \hspace{0.01cm} 2  & $\mathbf{j} \parallel Y$  \hspace{0.01cm} $\to$ \hspace{0.01cm} 1   \\
 \hline
 \hline
\end{tabular}
\end{table}

In order to characterize the $\mathbf{k-k'-j'}$ three-vector correlation, we make use of the alignment moments $A^{(2)}_0 (\mathbf{j'})$ and $A^{(2)}_{2+} (\mathbf{j'})$ of the diatomic's final rotational angular momentum $\mathbf{j'}$ with respect to the $XY$ plane. The $A^{(2)}_0 (\mathbf{j'})$ moment accounts for alignment of $\mathbf{j'}$ with respect to the initial relative velocity $\mathbf{k}$ and, in the high-$j'$ limit, ranges between $-1$ and $2$. Positive (negative) values of $A^{(2)}_0 (\mathbf{j'})$ correspond to $\mathbf{j' \parallel k}$ ($\mathbf{j' \perp k}$, in which case $\mathbf{j'}$  lies in the $XY$ plane). The $A^{(2)}_{2+} (\mathbf{j'})$ moment varies from $-1$ to $1$ (in the high-$j'$ limit). Its positive (negative) values correspond to alignment of $\mathbf{j'}$ along the $Y$-axis ($X$-axis). This is summarized in Table~\ref{table:moments}.

Figures~\ref{fig:A20_moment} and \ref{fig:A22_moment} display the $A^{(2)}_0 (\mathbf{j'})$ and $A^{(2)}_{2+} (\mathbf{j'})$ moments obtained in analytic form from the Fraunhofer model along with the results of experiment and close-coupling calculations of Wade \emph{et al.}~\cite{Wade04}. The agreement between the Fraunhofer model and the close-coupling calculation is compelling.

\begin{figure}
\includegraphics[width=8cm]{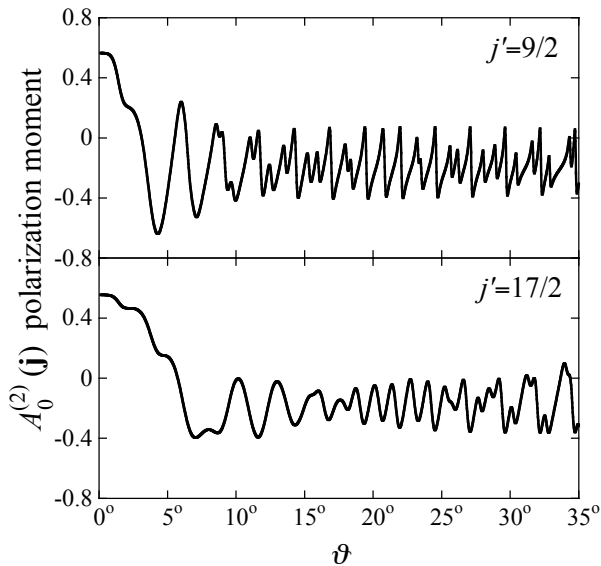}
\caption{\label{fig:k-j-kp} Polarization moment  $A^{(2)}_{0} (\mathbf{j})$ pertaining to the $\mathbf{k-j-k'}$ three-vector correlation in Ar--NO~$(j=\Omega=3/2, \to j', \Omega'=3/2)$ collisions at 520$\pm$70~cm$^{-1}$ obtained from the Fraunhofer model.}
\end{figure}
\begin{figure}

\includegraphics[width=8cm]{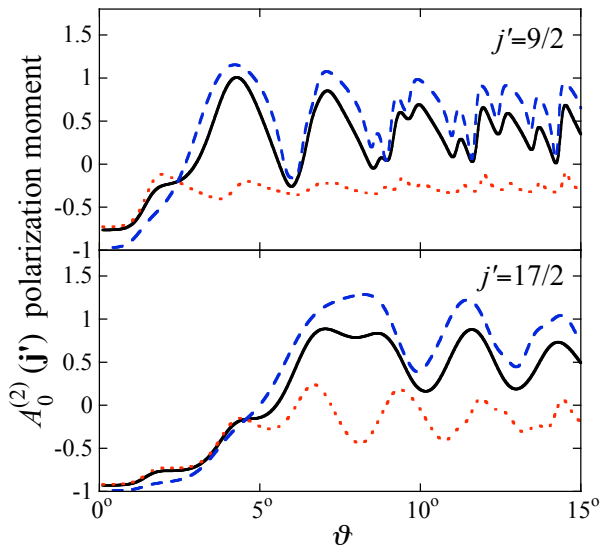}
\caption{\label{fig:k-j-kp-jp} The $\mathbf{k-j-k'-j'}$ four-vector correlation in the Ar--NO~$(j=\Omega=3/2, \to j', \Omega'=3/2)$ collisions at 520$\pm$70~cm$^{-1}$ in terms of the dependence of the final polarization moment $A^{(2)}_{0} (\mathbf{j'})$ on the initial alignment $A^{(2)}_{0} (\mathbf{j})=0$ (black line), $A^{(2)}_{0} (\mathbf{j})=-0.8$ (blue dashed line), and $A^{(2)}_{0} (\mathbf{j})=0.8$ (red dotted line). Obtained analytically from the Fraunhofer model.}
\end{figure}


Unfortunately, only two experimental points are available for the scattering angles concerned. Therefore, it is not clear whether the oscillatory behavior at small-angles would indeed show up in an experiment. We hope that the present work will inspire an experiment whose resolution will suffice to clarify this issue.

One can see that for zero scattering angle, $\vartheta=0$, $A^{(2)}_0 = -1$ and $A^{(2)}_{2+} = 0$. The reason is geometric: in pure forward scattering, the $\mathbf{j'}$ vector must be perpendicular to $\mathbf{k}$. Also, since $\mathbf{k}$ is roughly parallel to $\mathbf{k}'$ for very small $\vartheta$, the $\mathbf{j'}$ vector is approximately perpendicular to $\mathbf{k'}$. For small but nonzero scattering angles, $\vartheta \sim 5^\circ$, the $A^{(2)}_0$ moment becomes positive, both for $j'=9/2$ and $17/2$, indicating that $\mathbf{j'}$ tends to align along $\mathbf{k}$.

The $A^{(2)}_{2+}$ moment, on the other hand, exhibits narrow positive oscillations at very small angles ($\vartheta \approx 1^\circ$), but is in general negative, which corresponds to alignment of $\mathbf{j'}$ along the $X$-axis. 

Interestingly, the polarization moments presented in this paper are in a quantitative agreement with the accurate, close-coupling calculations, while the differential cross sections for the Ar--NO scattering, evaluated in ref.~\cite{LemFriFraunhofer}, agree only qualitatively. From this we draw the conclusion that the polarization moments are mainly due to the hard-core part of the potential. This conclusion is also supported by purely classical arguments based on the conservation of the projection of angular momentum on the collision's kinematic apse, see, e.g., ref.~\cite{Khare81}. The phase shift of the moment's oscillations as derived from the model with respect to those obtained from the close-coupling calculation is likely due to neglecting in the model the attractive part of the Ar--NO potential. This explanation is supported by a generalized Fraunhofer model that accounts for both attraction and repulsion~\cite{FullPotFraun} and which shifts the oscillations back toward smaller $\vartheta$.

The alignment moments of Figs.~\ref{fig:A20_moment} and \ref{fig:A22_moment} were normalized by the differential cross sections obtained from the Fraunhofer model. Since the Fraunhofer differential cross sections decrease faster with the scattering angle than their close-coupling counterparts~\cite{LemFriFraunhofer}, the oscillations of the analytic polarization moments are left relatively undamped at large scattering angles.

Since the $\mathbf{j}$ vector with $j=1/2$ can only be oriented but not aligned, the only-to-alignment-sensitive Fraunhofer model cannot handle vector correlations involving the $\mathbf{j}$-vector with $j=1/2$. However, since the $\mathbf{j}$ vector with $j=3/2$ can be aligned, we worked out the $\mathbf{k-j-k'}$ and  $\mathbf{k-j-k'-j'}$ vector correlations for the Ar -- NO~$(j=3/2,\Omega, \to j', \Omega)$ scattering within the $\Omega=3/2$ manifold. Figure~\ref{fig:k-j-kp} displays the  $A^{(2)}_{0} (\mathbf{j})$  polarization moment for the $\mathbf{k-j-k'}$ three-vector correlation. One can see that the small-angle scattering is favored by positive values of the $A^{(2)}_{0} (\mathbf{j})$ moment, which corresponds to $\mathbf{j \parallel k}$, i.e., to a ``broadside'' approach of NO with respect to $\mathbf{k}$, which enhances the scattering cross section. However, for larger scattering angles, the $A^{(2)}_{0} (\mathbf{j})$ moment becomes slightly negative, attesting to a preference for an ``edge-on'' approach with $\mathbf{j \perp k}$.

Figure~\ref{fig:k-j-kp-jp} exemplifies the $\mathbf{k-j-k'-j'}$ four-vector correlation in terms of the alignment moment $A^{(2)}_{0} (\mathbf{j'})$ of the final $\mathbf{j'}$ for different polarizations $A^{(2)}_{0} (\mathbf{j})$ of the initial $\mathbf{j}$. For an unpolarized initial state (black line), $\mathbf{j'}$ tends to align perpendicular to $\mathbf{k}$ (``broadside'' recoil) for very small $\vartheta$, but reverses to a slight alignment in the parallel direction (``edge-on'' recoil) for larger scattering angles. Initial polarization of NO such that $\mathbf{j \perp k}$ is seen to result in only small changes of the final alignment (blue dashed line). However, in the case of a ``broadside'' approach,  $\mathbf{j \parallel k}$, the stereodynamics changes significantly (red dotted line). The $A^{(2)}_{0} (\mathbf{j'})$ moment remains slightly negative throughout the range of scattering angles, indicating a propensity for an ``edge-on'' recoil, with  $\mathbf{j' \perp k}$.

The Fraunhofer model readily explains the above results: the analytic scattering amplitudes are proportional to the Bessel functions, which is a signature feature of diffraction. It is thus a diffractive oscillatory pattern that determines the angular dependence of the polarization moments. While the shape and frequency of the angular oscillations are entirely determined by the hard core of the PES, their position is somewhat influenced by the PES's attractive branch~\cite{FullPotFraun}. The Clebsch-Gordan coefficients that appear in the scattering amplitude bring about selection rules that constrain the final parity of the states and the projections of the angular momentum $\mathbf{j'}$ on $\mathbf{k}$. Within the model, the shape of the scatterer enters through the Legendre moments of a series expansion of the hard-core PES in terms of Legendre polynomials, $P_\kappa (\cos \theta)$. The angular momentum algebra that the model entails gives rise to additional selection rules which allow for nonzero contributions to the polarization moments to arise only from Legendre moments of order $\kappa \ge j'-j$. Therefore, the vector correlations for the $j=1/2,~3/2 \to j'=9/2$ channels are governed by the Legendre moment with $\kappa = 4$, whereas the $\kappa = 8$ Legendre moment governs the polarization moments of the $j=1/2,~3/2 \to j'=17/2$ channels.

Moreover, since the Fraunhofer model can account for collisions in an electrostatic field~\cite{LemFriFraunhofer}, we investigated the effect of the field on the polarization moments. A field of 16 kV/cm, sufficient to significantly orient the NO molecule in the space-fixed frame,  was found to cause only a tiny difference in the parity-resolved polarization moments as compared with the field-free ones. Upon averaging over the $e/f$ states, the effect of the field was found to be altogether negligible.

In summary, we made use of the Fraunhofer model of direct rotationally inelastic atom--diatom collisions to study vector correlations in such collisions analytically. The vector correlations obtained from the model closely reproduce those extracted from close-coupling calculations which, in turn, agree well with experiment. The Fraunhofer model of vector correlations demonstrates that the stereodynamics of the Ar--NO rotationally inelastic collisions is contained solely in the diffractive
part of the scattering amplitude which is governed by a single Legendre moment characterizing the anisotropy of the hard-core part of the system's PES. 
Given the ``geometric'' origin of this behavior -- ordained by the angular momentum algebra -- we expect to find a similar behavior in other systems.


We thank Marcelo de Miranda and Pablo Jambrina for helpful discussions, Elisabeth Wade and David Chandler for making available to us the results of their experiments and computations, and Millard Alexander for the Ar--NO PES. We are grateful to Gerard Meijer for discussions, encouragement, and support.


\end{document}